\PassOptionsToPackage{draft}{hyperref}
\documentclass[fleqn,usenatbib]{mnras}

\usepackage{newtxtext,newtxmath}
\usepackage{comment}
\usepackage{booktabs}
\usepackage[T1]{fontenc}
\usepackage{ae,aecompl}
\usepackage{siunitx}
\usepackage{cleveref}

\usepackage{graphicx}	% Including figure files
\usepackage{amsmath}	% Advanced maths commands
\usepackage{amssymb}	% Extra maths symbols

\usepackage{xspace}

\usepackage[xindy, toc, hyperfirst=false, nolist, nostyles, sanitize={name=false,description=false,symbol=false}]{glossaries}
\glsdisablehyper
\usepackage[hyperref,x11names, table]{xcolor}

\newcommand{\msun}{\hbox{$M_{\odot}$}}

\newcommand{\lsun}{\hbox{$L_{\odot}$}}

\newglossaryentry{vrad}{name={radial velocity~}, text={radial velocity}, symbol={\ensuremath{v_\textrm{rad}}}, description={radial velocity}, sort=vrad}
\newglossaryentry{vrot}{name={stellar rotation~}, name={stellar rotation}, symbol={\ensuremath{v_\textrm{rot}}}, description={radial velocity}, sort=vrot}

\newcommand{\kms}{\ensuremath{\textrm{km}~\textrm{s}^{-1}}}

\newcommand{\xray}{X-ray}

\newcommand{\nucl}[2]{\ensuremath{^{#2}\textrm{#1}}}
\newglossaryentry{angstrom}{name=\AA, description={unit of length $10^{-10}$\,m}, sort=angstrom}
\newglossaryentry{nir}{name=NIR,description={near infrared},first = {near infrared (NIR)}}
\newglossaryentry{psf}{name=PSF,description={point-spread function},first = {point-spread function (PSF)}}
\newglossaryentry{fwhm}{name=FWHM,description={Full Width Half Maximum},first = {FWHM}}
\newglossaryentry{rms}{name=RMS,description={Root Mean Square},first = {RMS}}
\newglossaryentry{signalnoise}{name=S/N,description={signal to noise}}
\newglossaryentry{uv}{name=UV,description={ultra violet},first = {ultra violet (UV)}}
\newglossaryentry{halpha}{name=\ensuremath{\textrm{H}\alpha}, description={First line of the Balmer series at 6563\,\AA}, sort=halpha}
\newglossaryentry{mgb}{name={Mg \textsc{i} b}, description={Triplet at 5167\,\AA, 5173\,\AA and 5184\,\AA}}
\newglossaryentry{sobolevapprox}{name={Sobolev approximation}, description={Lines are approximation with an infinitley thin interaction region \citep[e.g. no broadening][]{1960mes..book.....S}}, first={Sobolev approximation }}
\newglossaryentry{radeq}{name={radiative equilibrium}, description={The net flux of energy between matter and radiation field is zero}}
\newglossaryentry{nebularapprox}{name={nebular approximation}, description={Assumes that the plasma condition are controlled by a central radiation source. The radiation field decreases with the distance to the source by geometrical dilution. See \citet{1978stat.book.....M} for details}}
\newglossaryentry{modnebularapprox}{name={modified nebular approximation}, description={In contrast to \gls{nebularapprox} where only geometrical dilution is taken into account, the modified nebular approximation also takes dilution by other radiative processes into account }, first={modified nebular approximation}, parent=nebularapprox}
\newglossaryentry{thompsonscat}{name={Thomson scattering}, description={Scattering of photons on low energy electrons}}
\newglossaryentry{lte}{name={LTE}, description={Local Thermodynamic Equilibrium}, first={local thermodynamic equilibrium (LTE)}}
\newglossaryentry{lsr}{name={LSR}, description={Local Standard of Rest}, first={\textit{local standard of rest} (LSR)}}
\newglossaryentry{mc}{name={MC}, description={Monte Carlo}, first={\textit{Monte Carlo} (MC)}}
\newglossaryentry{wcs}{name={WCS}, description={world coordinate system}, first={world coordinate system (WCS)}}
\newglossaryentry{cmf}{name=CMF, text=CMF, first=Comoving Frame (CMF henceforth), description={Comoving Frame}}

\newglossaryentry{uvoir}{name=UVOIR, text=UVOIR, first=UV/optical/Near-IR (UVOIR), description={UV/optical/Near-IR}}

\newglossaryentry{sfit}{name=SFIT, text=\textsc{sfit}, description={spectral fitting program for hot stars \citep{2001A&A...376..497J}}, first={\textsc{sfit} \citep{2001A&A...376..497J}}}
\newglossaryentry{iraf}{name=IRAF, text=\textsc{iraf}, description={Image Reduction and Analysis Facility maintained by NOAO}, first={\textsc{iraf}\protect\footnote{IRAF: the Image Reduction and Analysis Facility is distributed by the National Optical Astronomy Observatory, which is operated by the Association of Universities for Research in Astronomy (AURA) under cooperative agreement with the National Science Foundation (NSF).}}}
\newglossaryentry{pyraf}{name=PyRAF, text=\textsc{PyRAF}, description={Python wrap of \gls{iraf} maintained by STSCI}, first=\textsc{PyRAF} \protect\footnote{PyRAF is a product of the Space Telescope Science Institute, which is operated by AURA for NASA.}}
\newglossaryentry{astropy}{name=ASTROPY, text=\textsc{astropy}, description=\textsc{astropy} framework, first = \textsc{astropy} \citep{2013A&A...558A..33A}}
\newglossaryentry{numpy}{name=NUMPY, text=\textsc{numpy}, description=\textsc{numpy} framework, first = \textsc{numpy} \citep{walt2011numpy}}
\newglossaryentry{scipy}{name=SCIPY, text=\textsc{scipy}, description=\textsc{scipy} framework, first = \textsc{scipy} \citep{Jones:2001fk}}
\newglossaryentry{matplotlib}{name=matplotlib, text=\textsc{matplotlib}, description=\textsc{matplotlib} framework, first = \textsc{matplotlib} \citep{hunter2007matplotlib}}
\newglossaryentry{pandas}{name=pandas, text=\textsc{pandas}, description=\textsc{pandas} framework, first = \textsc{pandas} \citep{mckinney2010data}}
\newglossaryentry{ipython}{name=ipython, text=\textsc{ipython}, description=\textsc{ipython} framework, first = \textsc{ipython} \citep{perez2007ipython}}
\newglossaryentry{jupyter}{name=jupyter, text=\textsc{jupyter}, description=\textsc{jupyter} framework, first = \textsc{jupyter} \citep{kluyver2016jupyter,perez2015project,ragan2014jupyter}}
\newglossaryentry{aplpy}{name=aplpy, text=\textsc{aplpy}, description=\textsc{aplpy} framework, first = \textsc{aplpy} \citep{2012ascl.soft08017R}}
\newglossaryentry{nltk}{name=nltk, text=\textsc{nltk}, description=\textsc{nltk} framework, first = Natural Language ToolKit \citep[\textsc{NLTK};][]{bird2009natural}}
\newglossaryentry{scikit-learn}{name=scikit-learn, text=\textsc{scikit-learn}, description=\textsc{scikit-learn} framework, first = \textsc{scikit-learn} \citep[][]{scikit-learn}}
\newglossaryentry{scikit-image}{name=scikit-image, text=\textsc{scikit-image}, description=\textsc{scikit-image} framework, first = \textsc{scikit-image} \citep[][]{scikit-image}}
\newglossaryentry{moog}{name=MOOG,text={\textsc{moog}}, description={spectral synthesis software \citep{1973ApJ...184..839S}}, first={\textsc{Moog} \citep{1973ApJ...184..839S}}}
\newglossaryentry{atlas9}{name=ATLAS9,description={grid of stellar atmospheres \citep{2004astro.ph..5087C}}, first={ATLAS9 \citep{2004astro.ph..5087C}}}
\newglossaryentry{vald}{name=VALD,description={Vienna Atomic Line Database \citep{2000BaltA...9..590K}}, first={Vienna Atomic Line Database \citep[VALD;][]{2000BaltA...9..590K}}}
\newglossaryentry{sextractor}{name=SExtractor, text=\textsc{SExtractor}, description={Source Extractor photometry program \citep{1996A&AS..117..393B}}, first={\textsc{SExtractor} \citep{1996A&AS..117..393B}}}

\newglossaryentry{swarp}{name=SWarp, text=\textsc{SWarp}, description={SWarp \citep{2002ASPC..281..228B}}, first={\textsc{SWarp} \citep{2002ASPC..281..228B}}}
\newglossaryentry{astrometry.net}{name=astrometry.net, text=\textsc{astrometry.net}, description={\textsc{astrometry.net} \citep{2010AJ....139.1782L}} first={\textsc{astrometry.net} \citep{2010AJ....139.1782L}}}

\newglossaryentry{astrodrizzle}{name=AstroDrizzle, text=\textsc{AstroDrizzle}, description={AstroDrizzle \citep{2012drzp.book.....G}}, first={\textsc{AstroDrizzle} \citep{2012drzp.book.....G}}}

\newglossaryentry{idl}{name=IDL,text={\textsc{idl}}, description={Interactive Data Language}}
\newglossaryentry{makee}{name=MAKEE,text=\textsc{makee}, description={MAuna Kea Echelle Extraction by Tom Barlow available}}% at \verb+http://spider.ipac.caltech.edu/staff/tab/makee/index.html+}}
\newglossaryentry{minuit}{name=MINUIT,text={\textsc{minuit}}, description={collection of numerical optimization tools \citep{James:1975dr}}}
\newglossaryentry{migrad}{name=MIGRAD,text={\textsc{migrad}}, description={numerical gradient optimization tools - part of \gls{minuit}}}
\newglossaryentry{dolphot}{name=DOLPHOT, text=\textsc{dolphot}, description=photometry package for HST, first=\textsc{dolphot} \citep{2000PASP..112.1383D}}
\newglossaryentry{synphot}{name=synphot, text={\textsc{synphot}}, description={synthetic photometry package from STSCI}, first={\textsc{synphot}\protect\footnote{\textsc{synphot} is a product of the Space Telescope Science Institute, which is operated by AURA for NASA.}}}
\newglossaryentry{chianti}{name=CHIANTI, text=CHIANTI, description= CHIANTI Database 7.1, first =CHIANTI 7.1 \citep{1997A&AS..125..149D,2012ApJ...744...99L}}
\newglossaryentry{synpp}{name=SYNPP, text=SYN++, description= SYN++ software, first =SYN++ \citep{2011PASP..123..237T}}
\newglossaryentry{tardis}{name=TARDIS, text=\textsc{tardis}, description= TARDIS MC code, first = {\textsc{tardis} \citep{2014MNRAS.440..387K}}}

\newglossaryentry{artis}{name=ARTIS, text=\textsc{artis}, description= ARTIS MC code, first = \textsc{artis} \citep{2009MNRAS.398.1809K}}
\newglossaryentry{cmfgen}{name=CMFGEN, text=\textsc{cmfgen}, description=CMFGGEn radiative transfer code, first = \textsc{cmfgen} \citep{1998ApJ...496..407H}}
\newglossaryentry{sedona}{name=SEDONA, text=\textsc{sedona}, description= Sedona MC code, first = \textsc{sedona} \citep{2006ApJ...651..366K}}
\newglossaryentry{mlmc}{name=MLMC, text=ML93, description= Mazzali Lucy Monte Carlo, first ={Mazzali \& Lucy (1993, ML93) code}}
\newglossaryentry{starkit}{name=STARKIT, text=\textsc{starkit}, description= TARDIS MC code, first = {\textsc{starkit} \citep{wolfgang_kerzendorf_2015_28016}}}

\newglossaryentry{pyne}{name=PYNE, text=\textsc{pyne}, description= PYNE code, first = {\textsc{pyne} \citep{Scopatz2012a}}}
\newglossaryentry{multinest}{name=MULTINEST, text=\textsc{MultiNest}, description=MultiNest, first={\textsc{MultiNest} \citep{2009MNRAS.398.1601F}}}
\newglossaryentry{wsynphot}{name=WSYNPHOT, text=\textsc{wsynphot}, description=Wsynphot, first={\textsc{wsynphot}\protect\footnote{\url{https://github.com/wkerzendorf/wsynphot}}}}
\newglossaryentry{specutils}{name=SPECUTILS, text=\textsc{specutils}, description=specutils, first={\textsc{specutils} \protect\footnote{\url{https://github.com/astropy/specutils}}}}
\newglossaryentry{ads}{name=ADS ,description=ADS, first={NASA Astrophysics Data System (ADS) \citep{2000A&AS..143...41K}}}

\newglossaryentry{2mass}{name=2MASS,description={Two Micron All Sky Survey \citep{2006AJ....131.1163S}}, first={Two Micron All Sky Survey \citep{2006AJ....131.1163S}}}
\newglossaryentry{nomad}{name=NOMAD,first={Naval Observatory Merged Astrometric Dataset \citep[NOMAD; ][]{2005yCat.1297....0Z}}, description={Naval Observatory Merged Astrometric Dataset}}
\newglossaryentry{wifes}{name=WIFES, text=\textsc{WiFeS}, first={\textsc{WiFeS} \citep{2007Ap&SS.310..255D}},  description={Wide Field Spectrograph - \gls{ifu} mounted on the 2.3\,m telescope at Siding Spring Observatory}}
\newglossaryentry{scp}{name=SCP,description={Supernova Cosmology Project, led by Saul Perlmutter}, first={Supernova Cosmology Project (SCP)}}
\newglossaryentry{hzsns}{name=HZSNS,description={High Z Supernova Search, led by Brian Schmidt}, first={High Z Supernova Search (HZSNS)}}
\newglossaryentry{vlt}{name=VLT,description={Very Large Telescope located on Cerro Paranal (Chile)}, first={Very Large Telescope (VLT)}}
\newglossaryentry{flames}{name=FLAMES,description={Multi-object, intermediate and high resolution spectrograph mounted on the  \gls{vlt}}}
\newglossaryentry{hires}{name=HIRES, description={High Resolution Echelle Spectrometer mounted on the Keck Telescope}, first={High Resolution Echelle Spectrometer \citep[HIRES;][]{1994SPIE.2198..362V}}}
\newglossaryentry{lris}{name=LRIS,description={Low Resolution Imaging Spectrometer mounted on the Keck Telescope}, first={Low-Resolution Imaging Spectrometer \citep[LRIS;][]{Oke95}}}

\newglossaryentry{decam}{name=DECam, description={DECam is a high-performance, wide-field CCD imager mounted at the prime focus of the Blanco 4-m telescope at \gls{ctio}.}, first={Dark Energy Camera \citep[DECam; ][]{2012PhPro..37.1332D,2015AJ....150..150F}}}

\newglossaryentry{essence}{name=ESSENCE,description={The `Equation of State: SupErNovae trace Cosmic Expansion' project \citep[ESSENCE;][]{2002AAS...201.7809G}}, first={`The Equation of State: SupErNovae trace Cosmic Expansion' \citep[ESSENCE;][]{2002AAS...201.7809G}}}
\newglossaryentry{ifu}{name=IFU,description={Optical instrument combining spectrographic and imaging capabilities, used to obtain spatially resolved spectra}, first={Integral Field Unit (IFU)}, firstplural={Integral Field Units (IFUs)}}

\newglossaryentry{besancon}{name=Besan\c{c}on Model, description={Model of stellar population synthesis of the Galaxy, including kinematics.}}%  \verb+http://model.obs-besancon.fr+} }, nonumberlist=true}

\newglossaryentry{int}{name=INT,description={Isaac Newton 2.5\,m Telescope}, first={Isaac Newton 2.5\,m Telescope (INT)}}
\newglossaryentry{iau}{name=IAU,description={International Astronomical Union}, first={IAU}}
\newglossaryentry{chandra}{name=Chandra,description={Chandra \xray\ Observatory (space-based)}}
\newglossaryentry{hst}{name=HST,description={Hubble Space Telescope}}
\newglossaryentry{hst.wfpc2}{name=WFPC2,description={Wide-Field Planetary Camera 2 mounted on the \gls{hst}}, first={Wide-Field Planetary Camera 2 (WFPC2)}}
\newglossaryentry{hst.acs}{name=ACS,description={Advanced Camera for Surveys mounted on the \gls{hst}}, first={Advanced Camera for Surveys (ACS)}}
\newglossaryentry{hst.wfc3}{name=WFC3,description={Wide-Field Camera 3 mounted on the \gls{hst}}, first={Wide-Field Camera 3 (WFC3)}}
\newglossaryentry{hst.cte}{name=CTE, description={charge transfer efficiency (CTE)}, first={charge transfer efficiency \citep[CTE; see ][for a description]{2009acs..rept....1C}}}

\newglossaryentry{snls}{name=SNLS,description={Supernova Legacy Survey \citep{2003AAS...203.8209P}}, first={Supernova Legacy Survey \citep[SNLS;][]{2003AAS...203.8209P}}}
\newglossaryentry{dass}{name=DASS, description={Digitized Astronomy Supernova Survey \citep{1975PASP...87..565C}}, first={Digitized Astronomy Supernova Survey \citep[DASS;][]{1975PASP...87..565C}}}
\newglossaryentry{bait}{name=BAIT, description={Berkley Automatic Imaging Telescope \citep{1993PASP..105.1164R}}, first={Berkley Automatic Imaging Telescope \citep[BAIT;][]{1993PASP..105.1164R}}}
\newglossaryentry{kait}{name=KAIT, description={Katzman Automatic Imaging Telescope \citep{2001ASPC..246..121F}}, first={Katzman Automatic Imaging Telescope \citep[KAIT;][]{2001ASPC..246..121F}}}
\newglossaryentry{loss}{name=LOSS, description={Lick Observatory Supernova Search  \citep{2000AIPC..522..103L}}, first={Lick Observatory Supernova Search \citep[LOSS;][]{2000AIPC..522..103L}}}
\newglossaryentry{ctss}{name=CTSS,description={Cal\'{a}n/Tololo Supernova Survey \citep{1993AJ....106.2392H}}, first={Cal\'{a}n/Tololo supernova survey \citep[CTSS;][]{1993AJ....106.2392H}}}
\newglossaryentry{ctio}{name= CTIO, description={Cerro Tololo Inter-American Observatory}, first={Cerro Tololo Inter-American Observatory (CTIO)}}
\newglossaryentry{ptf}{name=PTF, description={Palomar Transient Factory \citep{2009PASP..121.1334R}}, first={Palomar Transient Factory \citep[PTF;][]{2009PASP..121.1334R}}}
\newglossaryentry{batse}{name=BATSE, description={Burst and Transient Source Experiment mounted on the Compton Gamma Ray Observatory}, first={Burst and Transient Source Experiment (BATSE)}}
\newglossaryentry{bepposax}{name=BeppoSAX, description={\xray\ satellite named in honor of Giuseppe "Beppo" Occhialini}}
\newglossaryentry{rosat}{name=ROSAT, description={short for R\"{o}ntgensatellit}, first={ROSAT}}
\newglossaryentry{hete2}{name=HETE2, description={High Energy Transient Explorer}, first={High Energy Transient Explorer (HETE)}}
\newglossaryentry{ska}{name=SKA, description={Square Kilometre Array}, first={Square Kilometre Array (SKA)}}

\newglossaryentry{gnirs}{name=GNIRS, description={Gemini Near InfraRed Spectrograph mounted on the Gemini North Telescope}}
\newglossaryentry{gmosn}{name=GMOS, description={Gemini Multi Object Spectrograph mounted on the
 Gemini North Telescope}, first={GMOS \citep[Gemini Multi Object Spectrograph;][]{2004PASP..116..425H}}}
\newglossaryentry{swift}{name=Swift, description={Swift Gamma-Ray Burst Mission}}
\newglossaryentry{vla}{name=VLA, description={Very Large Array radio telescope located in North America}, first={Very Large Array (VLA)}}
\newglossaryentry{evla}{name=EVLA, description={Extended Very Large Array radio telescope located in North America}, first={Extended Very Large Array (EVLA)}}
\newglossaryentry{sdss}{name=SDSS, description={Sloan Digital Sky Survey}}
\newglossaryentry{dss}{name=DSS, description={Digitized Sky Survey}}
\newglossaryentry{skymapper}{name=SkyMapper, description={SkyMapper telescope \citep{2007PASA...24....1K}}, first={SkyMapper \citep{2007PASA...24....1K}}}
\newglossaryentry{panstarrs}{name=PanSTARRS, description={Panoramic Survey Telescope \& Rapid Response System \citep{2004SPIE.5489...11K}}, first={Panoramic Survey Telescope \& Rapid Response System \citep[PanSTARRS;][]{2004SPIE.5489...11K}}}
\newglossaryentry{lsst}{name=LSST, description={Large Synoptic Survey Telescope}, first={Large Synoptic Survey Telescope \citep[LSST;][]{2006AAS...209.8604P}}}
\newglossaryentry{ppmxl}{name=PPMXL, description={PPMXL Catalog of Positions and Proper Motions on the ICRS \citep{2010AJ....139.2440R}}}
\newglossaryentry{gaia}{name=GAIA, description={Global Astrometric Interferometer for Astrophysics \citep{2001A&A...369..339P}}, first={Global Astrometric Interferometer for Astrophysics \citep[GAIA;][]{2001A&A...369..339P}}}
\newglossaryentry{ligo}{name=LIGO, description={Laser Interferometer Gravitational Wave Observatory}, first={Laser Interferometer Gravitational Wave Observatory \citep[LIGO;][]{1992Sci...256..325A}}}
\newglossaryentry{aligo}{name=Advanced LIGO, description={Advanced LIGO}, sort=ligo2}
\newglossaryentry{lisa}{name=LISA, description={Laser Interferometer Space Antenna \citep{1994ESAJ...18..219J}}, first={Laser Interferometer Space Antenna \citep[LISA;][]{1994ESAJ...18..219J}}}
\newglossaryentry{ccd}{name=CCD,description={Charged Coupled Device}, first={charged coupled device (CCD)}, firstplural={charged coupled devices (CCDs)}}

\newcommand{\sn}[2]{SN~#1#2\xspace}

\newglossaryentry{irc}{name=IRC, text={IRC}, description={infrared catastrophe}, first={infrared catastrophe \citep[IRC;][]{1980PhDT.........1A}}}

\newglossaryentry{sn}{name=Supernova, text={SN}, plural={SNe}, description={exploding star}, nonumberlist=true, first={supernova (SN)}, firstplural={supernovae (SNe)}}
\newglossaryentry{snia}{name=Type~Ia (SN~Ia), text={SN~Ia}, description={Thermonuclear explosion of a white dwarf - spectra show no hydrogen but a strong silicon line},first={Type~Ia supernova (SN~Ia)}, firstplural={Type Ia supernovae (SNe~Ia)}, plural={SNe~Ia}, parent=sn, nonumberlist=true}
\newcommand{\sneia}{\glspl*{snia}\xspace}
\newcommand{\snia}{\gls*{snia}\xspace}

\newglossaryentry{branchnormal}{name={branch-normal}, text=\textit{Branch-normal}, description={Large homogeneous class of Type Ia Supernovae, defined in \citet{1993AJ....106.2383B}}, first={\textit{Branch-normal} SNe Ia \citep{1993AJ....106.2383B}}, parent=snia}
\newglossaryentry{91t}{name={91T-like}, description={Luminous class of Type Ia supernovae similar to \sn{1991}{T} \citep{1992AJ....103.1632P}} , first={91T-like}, parent=snia}
\newglossaryentry{91bg}{name={91bg-like}, description={Faint class of Type Ia supernovae similar to \sn{1991}{bg} \citep{1992AJ....104.1543F}}, first={91bg-like}, parent=snia}
\newglossaryentry{02cx}{name={02cx-like}, description={Peculiar class of Type Ia supernovae similar to \sn{2002}{cx} \citep{2003PASP..115..453L}}, first={02cx-like \sneia\ \citep{2003PASP..115..453L}}, parent=snia}

\newglossaryentry{snibc}{name=Type~Ib/c, text={SN~Ib/c}, description={Collapse of the core of a massive star -  spectrum shows no hydrogen and no silicon line},first={Type~Ib/c supernova (SN~Ib/c)}, firstplural={Type~Ib/c supernovae (SNe~Ib/c)}, plural={SNe~Ib/c}, parent=sn}

\newglossaryentry{snib}{name=Type~Ib, text={SN~Ib}, description={Spectrum shows no hydrogen and no silicon, but helium line},first={Type Ib supernova (SN~Ib)}, firstplural={Type~Ib supernovae (SNe~Ib)}, plural={SNe~Ib}, parent=snibc}

\newglossaryentry{snic}{name=Type~Ic, text={SN~Ic}, description={Spectrum shows no hydrogen, no silicon and no helium line},first={Type~Ic supernova (SN~Ic)}, firstplural={Type~Ic supernovae (SNe~Ic)}, plural={SNe~Ic}, parent=snibc}

\newglossaryentry{snii}{name=Type~II, text={SN~II}, description={Collapse of the core of a massive star - spectrum shows strong hydrogen line},first={Type~II supernova (SN~II)}, firstplural={Type~II supernovae (SNe~II)}, plural={SNe~II}, parent=sn}

\newglossaryentry{sniib}{name=Type~IIb, text={SN~IIb}, description={Spectrum shows hydrogen and helium lines},first={Type~IIb supernova (SN~IIb)}, firstplural={Type~IIb supernovae (SNe~IIb)}, plural={SNe~IIb}, parent=snii}

\newglossaryentry{sniip}{name=Type~II~Plateau (Type IIP), text={SN~IIP}, description={Lightcurve shows plateau},first={Type~IIP supernova (SN~IIP)}, firstplural={Type~II Plateau supernovae \citep[SNe~IIP;][]{1979A&A....72..287B}}, plural={SNe~IIP}, parent=snii}

\newglossaryentry{sniil}{name=SN~II~Linear, text={SN~IIL}, description={Lightcurve shows no plateau, but linear decline},first={Type~IIL supernova (SN~IIL)}, firstplural={Type~II~Linear supernovae \citep[SNe~IIL;][]{1990MNRAS.244..269S}}, plural={SNe~IIL}, parent=snii}

\newglossaryentry{sniin}{name=Type II narrow-lined (Type IIn), description={Spectrum shows narrow lines},first={Type~II~narrow-lined supernova (SN IIn)}, firstplural={Type~IIn supernovae (SNe~IIn)}, plural={SNe~IIn}, parent=snii}

\newglossaryentry{snr}{name=Remnant (SNR), text=SNR, description={Remnant left visible post-explosion}, first={supernova remnant (SNR)}, firstplural={supernova remnants (SNRs)}, parent=sn}

\newglossaryentry{dtd}{name=DTD,description={delay time distribution - expected supernova rate over time after a brief outburst of starformation},first={delay time distribution (DTD)}, firstplural={delay time distributions (DTDs)}, plural=DTDs}

\newglossaryentry{hvg}{name=HVG,description={high velocity gradient - Type Ia supernovae with a fast evolution of photospheric velocity},first={high velocity group (HVG)}, firstplural={high velocity groups (HVGs)}, plural=HVGs, parent=snia}

\newglossaryentry{lvg}{name=LVG,description={low velocity gradient - Type Ia supernovae with a slow evolution of photospheric velocity},first={low velocity group (LVG)}, firstplural={low velocity groups (LVGs)}, plural=LVGs, parent=snia}

\newglossaryentry{wd}{name=white dwarf (WD), text=WD, description={White Dwarf - extremely dense stellar remnant}, first={white dwarf (WD)}}
\newglossaryentry{onemgwd}{name= Oxygen/Neon (ONe), text={ONe-WD},description={Oxygen/Neon White Dwarf}, first={oxygen/neon White Dwarf (ONe-WD)}, parent=wd}
\newglossaryentry{cowd}{name=carbon/oxygen (CO), text={CO-WD}, description={carbon/oxygen white dwarf}, first={carbon/oxygen white dwarf (CO-WD)}, firstplural = {carbon/oxygen white dwarfs (CO-WDs)}, parent=wd}

\newglossaryentry{sds}{name=SD-Scenario,description={single-degenerate scenario (single white dwarf accreting from non-degenerate companion)}, first={single-degenerate scenario (SD-scenario)}}

\newglossaryentry{dds}{name=DD-Scenario, description={double degenerate scenario (merging of two white dwarfs)}, first={double-degenerate scenario (DD-scenario)}}

\newglossaryentry{sss}{name=SSS, text={supersoft \xray\ source}, description={supersoft \xray\ source - believed to be emitted by nuclear fusion on a white dwarf's surface}}%, first={supersoft \xray\ source (SSS)}, firstplural={supersoft \xray\ sources (SSS)}}

\newglossaryentry{amcvn}{name=AM CVn, description={AM Canum Venaticorum star \citep[white dwarf accreting hydrogen poor matter from a companion star; see ][]{2005ASPC..330...27N}}}

\newglossaryentry{rlof}{name=RLOF, description={Roche Lobe Overflow (see \citet{1971ARA&A...9..183P} for a more detailed description)}, first={Roche-lobe overflow (RLOF)}}

\newglossaryentry{mchan}{name={Chandrasekhar mass~}, text={Chandrasekhar~mass}, symbol={\ensuremath{M_\textrm{Chan}}}, plural={Chandrasekhar~masses}, description={Mass when the core of a star collapses due to insufficient degeneracy pressure - for a white dwarf $\approx1.38\,M_\odot$ see \citet{1931ApJ....74...81C}}, first={Chandrasekhar~mass \citep[$M_\textrm{Chan}=1.38\,M_\odot$;][]{1931ApJ....74...81C}}, sort=mchan}

\newglossaryentry{w7}{name={W7 model},description={W7 model \citep{1984ApJ...286..644N}},first = {W7 model \citep{1984ApJ...286..644N}}}

\newglossaryentry{ew}{name=Equivalent Width, text={EW}, description={width of a rectangle that has the same area as a spectral line when taken to zero flux}, first={equivalent width (EW)}, firstplural={equivalent widths (EWs)}}
\newglossaryentry{agb}{name=AGB,description={Asymptotic Giant Branch}, first={Asymptotic Giant Branch (AGB)}}
\newglossaryentry{cmb}{name=CMB,description={Cosmic Microwave Background}}
\newglossaryentry{csm}{name=CSM,description={Circumstellar Medium}, first={circumstellar medium (CSM)}}
\newglossaryentry{csi}{name=CSI,description={Circumstellar Interaction}, first={circumstellar interaction (CSI)}}
\newglossaryentry{ism}{name=ISM,description={Interstellar Medium}, first={interstellar medium (ISM)}}
\newglossaryentry{ige}{name=IGE,description={Iron Group Element}, first={iron group element (IGE)}, firstplural={iron group elements (IGEs)}}
\newglossaryentry{epm}{name=EPM,description={Expanding Photosphere Method \citep{1974ApJ...193...27K}}, first={Expanding Photosphere Method (EPM)}}
\newglossaryentry{aic}{name=AIC,description={Accretion Induced Collapse}, first={accretion induced collapse (AIC)}}
\newglossaryentry{ime}{name=IME,description={Intermediate Mass Element}, first={intermediate mass element (IME)}, firstplural={intermediate mass elements (IMEs)}}
\newglossaryentry{h0}{name=\ensuremath{H_0},description={Hubbles constant}}
\newglossaryentry{nse}{name=NSE,description={Nuclear Statistical Equilibrium}, first={nuclear statistical equilibrium (NSE)}}
\newglossaryentry{cdm}{name=CDM,description={Cold Dark Matter}}
\newglossaryentry{grb}{name=GRB,description={Gamma Ray Burst}, first={Gamma Ray Burst (GRB)}, firstplural={Gamma Ray Bursts (GRBs)}}
\newglossaryentry{xps}{name=XPS, description={x-ray point source}, first={x-ray point source (XPS)}, firstplural={x-ray point sources (XPS)}}
\newglossaryentry{donor}{name=donor,description={non-degenerate companion in the \gls{sds}}}
\newglossaryentry{mainsequence}{name=main sequence,description={main sequence star}}
\newglossaryentry{redgiant}{name=red giant,description={red giant star}}
\newglossaryentry{mlcs}{name=MLCS,description={Multicolor Light Curve Shape method \citep[MLCS;][]{1996ApJ...473...88R}}, first={Multicolor Light-Curve Shape method \citep[MLCS;][]{1996ApJ...473...88R}}}
\newglossaryentry{rsoph}{name=RS~Ophiuci ,description={white dwarf accreting from a red giant - assumed progenitor of the \gls{sds}}, sort=rsoph}
\newglossaryentry{usco}{name=U~Scorpii,description={white dwarf accreting from a main sequence star - assumed progenitor of the \gls{sds}}, sort=usco}
\newglossaryentry{rcw86}{name=RCW~86,description={supernova remnant sometimes associated with \sn{185}{}}, sort=rcw86}
\newglossaryentry{casa}{name=Cas~A,description={Cassiopeia A supernova remnant - probably a \gls{snib} event}}
\newglossaryentry{cepheid}{name=Cepheid,description={very luminous variable star with a strong luminosity period relationship}}
\newglossaryentry{urca}{name=Urca, text=\textit{Urca}, description={process predominatly contributing to cooling in stars. The \textit{Urca} process consists of alternating electron-capture and $\beta^{-}$ decay of two nuclei pairs.},sort=urca}
\newglossaryentry{alphacen}{name=Alpha Centauri,description={one of the brightest stars in the night sky and a close binary}}
\newglossaryentry{pcygni}{name={P Cygni}, text={P Cygni},description={a hypergiant luminous blue variable with strong winds. Often referred to as a description for their line profiles showing a emission peak at the rest wavelength of the line and a blue-shifted absorption trough.}}

\newglossaryentry{teff}{name={effective temperature~}, text={effective temperature}, symbol={\ensuremath{T_\textrm{eff}}}, description={Temperature of a blackbody emitting the same total energy}, sort=teff}

\newglossaryentry{logg}{name={surface gravity~}, text={surface gravity}, symbol={\ensuremath{\textrm{log}\,g}}, description={gravity at the surface of a star}, sort=logg}
\newglossaryentry{feh}{name={metallicity~}, text={metallicity}, symbol=\textrm{[Fe/H]},description={iron abundance relative to the sun}, sort=feh}

\newglossaryentry{texp}{name={time since explosion~}, text={time since explosion}, text={time since explosion}, symbol={\ensuremath{t_{\rm exp}}},description={time since explosion (measured in days)}, sort=texp, first={time since explosion (\ensuremath{t_{\rm exp}})}}

\newglossaryentry{lmc}{name=LMC,description={Large Magellanic Cloud}, first={Large Magellanic Cloud (LMC)}, sort=lmc}
\newglossaryentry{smc}{name=SMC,description={Small Magellanic Cloud}, sort=smc}
\newglossaryentry{z}{name=\ensuremath{z},description={redshift}, sort=z}

\title[Search for a WD in SN~1006]{A Search for a Surviving White Dwarf Companion in SN~1006}

\author[Kerzendorf, W.E.~et al.]{
W.E.~Kerzendorf,$^{\!1, 2}$\thanks{E-mail: wkerzendorf@gmail.com} G.~Strampelli,$^{\!3}$ K.~J.~Shen,$\!^{4}$ J.~Schwab,$^{\!5, 6}$ R.~Pakmor,$^{\!7}$\newauthor
T.~Do,$^{\! 8}$  J. Buchner,$^{\!9,10}$ A.~Rest$^{3}$
\\
$^{1}${European Southern Observatory, Karl-Schwarzschild-Stra{\ss}e 2, 85748 Garching bei M\"{u}nchen, Germany}\\
$^{2}${ESO Fellow}\\
$^{3}${Space Telescope Science Institute AURA, 3700 San Martin Drive, Baltimore, MD 21218, USA}\\
$^{4}${Department of Astronomy and Theoretical Astrophysics Center, 
University of California, Berkeley, CA 94720, USA}\\
$^{5}${Department of Astronomy and Astrophysics, University of California, Santa Cruz, CA 95064, USA} \\
$^{6}${Hubble Fellow} \\
$^{7}${Heidelberger Institut f\"{u}r Theoretische Studien, Schloss-Wolfsbrunnenweg 35, D-69118 Heidelberg, Germany}\\
$^{8}${UCLA Galactic Center Group, Physics and Astronomy Department, UCLA, Los Angeles, CA 90095-1547, USA}\\
$^{9}${Millenium Institute of Astrophysics, Vicuna MacKenna 4860, 7820436 Macul, Santiago, Chile}\\
$^{10}${Pontificia Universidad Catlica de Chile, Instituto de Astrofisica, Vicuna Mackenna 4860, 7820436 Macul, Santiago, Chile}
}

\pubyear{2017}

\begin{document}
\label{firstpage}
\pagerange{\pageref{firstpage}--\pageref{lastpage}}
\maketitle

% Abstract of the paper
\begin{abstract}
Multiple channels have been proposed to produce Type Ia supernovae, with many scenarios suggesting that the exploding white dwarf accretes from a binary companion pre-explosion. In almost all cases, theory suggests that this companion will survive. However, no such companion has been unambiguously identified in ancient supernova remnants -- possibly falsifying the accretion scenario. Existing surveys, however, have only looked for stars as faint as $\approx 0.1\lsun$ and thus might have missed a surviving white dwarf companion. In this work, we present very deep DECAM imaging (\textit{u, g, r, z}) of the Type Ia supernova remnant \sn{1006}{} specifically to search for a potential surviving white dwarf companion. We find no object within the inner third of the \sn{1006}{} remnant that is consistent with a relatively young cooling white dwarf. We find that if there is a companion white dwarf, it must have formed long ago and cooled undisturbed for $> 10^8$ yr to be consistent with the redder objects in our sample. We conclude that our findings are consistent with the complete destruction of the secondary (such as in a merger) or an unexpectedly cool and thus very dim surviving companion white dwarf.

\end{abstract}

% Select between one and six entries from the list of approved keywords.
% Don't make up new ones.

%%%%%%%%%%%%%%%%%%%%%%%%%%%%%%%%%%%%%%%%%%%%%%%%%%

%%%%%%%%%%%%%%%%% BODY OF PAPER %%%%%%%%%%%%%%%%%%

\section{Introduction}

\sneia are explosions lacking hydrogen and producing copious amounts of \nucl{Ni}{56}. This behavior is consistent with thermonuclear burning in a massive ($ \geq 0.9 \msun$) degenerate \gls{wd}. There is no known mechanism for these objects to self-ignite and so theories of the origin of the explosion suggest that the \gls{wd} can be ignited through binary interaction. For this work, we distinguish between two classes of binary interaction. In one, the \gls{wd} fully merges with its companion prior to the explosion so that no star remains after the \snia occurs.  In the other, the companion remains separate from the exploding \gls{wd} and survives the explosion.

Detection of a surviving companion star post-explosion would serve to distinguish between the two scenarios \citep[e.g. ][]{2000ApJS..128..615M, 2008A&A...489..943P,2013ApJ...773...49P, 2013ApJ...765..150S}. Several groups have attempted to unambiguously identify such a donor star to no avail \citep[e.g.][]{2004ApJ...612..357R,2009ApJ...691....1G,2009ApJ...701.1665K,2014ApJ...782...27K, 2012Natur.481..164S}. However, these searches have all focused on relatively bright companions $>0.1\lsun$, so they could have missed a very faint surviving companion  \citep[e.g. a \gls{wd}; see][]{2017ApJ...834..180S}. 

Several theories suggest the possibility that the surviving companion is another \gls{wd}.  In double
degenerate double detonation scenarios, helium from a He or low-mass
C/O WD detonates on the surface of the more massive C/O WD, leading to
a carbon core detonation \citep{2007A&A...476.1133F,2014ApJ...785...61S}.  When this He is transferred stably, as in
AM Canum Venaticorum binaries \citep{2007ApJ...662L..95B}, or if the He shell detonates quickly
enough after the onset of unstable mass transfer, as in double WD
mergers \citep{2010ApJ...709L..64G,2013ApJ...770L...8P}, the less massive companion WD may survive the explosion of the
more massive WD.  In the single degenerate spin-up/spin-down scenario
\citep{2011ApJ...730L..34J, 2011ApJ...738L...1D}, a C/O WD accretes
material from a non-degenerate donor.  The accreted angular momentum spins
the WD up, allowing it to grow above the canonical (non-rotating) critical
mass needed for explosion.  However, the accretor WD does not explode
until it has spun-down, and during this time the donor star can evolve to
become a WD.

Deep photometric searches for surviving companions are difficult because most supernova remnants are located very close to the Galactic Plane \citep[mean and standard deviation of Galactic latitude of remnants $b=0.07\pm2.72\deg$; data from][]{2014BASI...42...47G} and thus heavily affected by extinction. Fortunately, one of the closest \citep[]{2003ApJ...585..324W}  and youngest remnants in the Galaxy, \sn{1006}{} lies very far above the plane ($b=14.6\deg$) and is only very mildly affected by dust \citep[$A_V\approx0.3$;][]{2011ApJ...737..103S}. Thus, \sn{1006}{} is the most promising candidate for a deep search for a possible surviving \gls{wd}. While previous observations of this remnant \citep[][]{2012ApJ...759....7K,2012Natur.489..533G} have not found any surviving companion, they did not probe deep enough to find a possible surviving \gls{wd}. 

%({\bf JWS: Do we need to cite some study for the identification of this as a Ia remnant?})

In this work, we present the deepest photometric study so far of \sn{1006}{} using \gls{decam} data and compare the resulting photometry with surviving \gls{wd} models by various authors. 

In Section~\ref{sec:observation}, we present the observations as well as our photometric measurement techniques. In Section~\ref{sec:analysis}, we present our techniques to compare the photometry with the suggested surviving \gls{wd} models. Section~\ref{sec:results} confronts the different theoretical scenarios with the analysis performed in this work. We conclude this work in Section~\ref{sec:conclusion} and give an overview of future possible tests of the scenarios.

\section{Observations \& Data reduction}
\label{sec:observation}
\begin{figure}
	% To include a figure from a file named example.*
	% Allowable file formats are eps or ps if compiling using latex
	% or pdf, png, jpg if compiling using pdflatex
	\includegraphics[width=\columnwidth,clip,trim={0 0 1.5cm 1.5cm}]{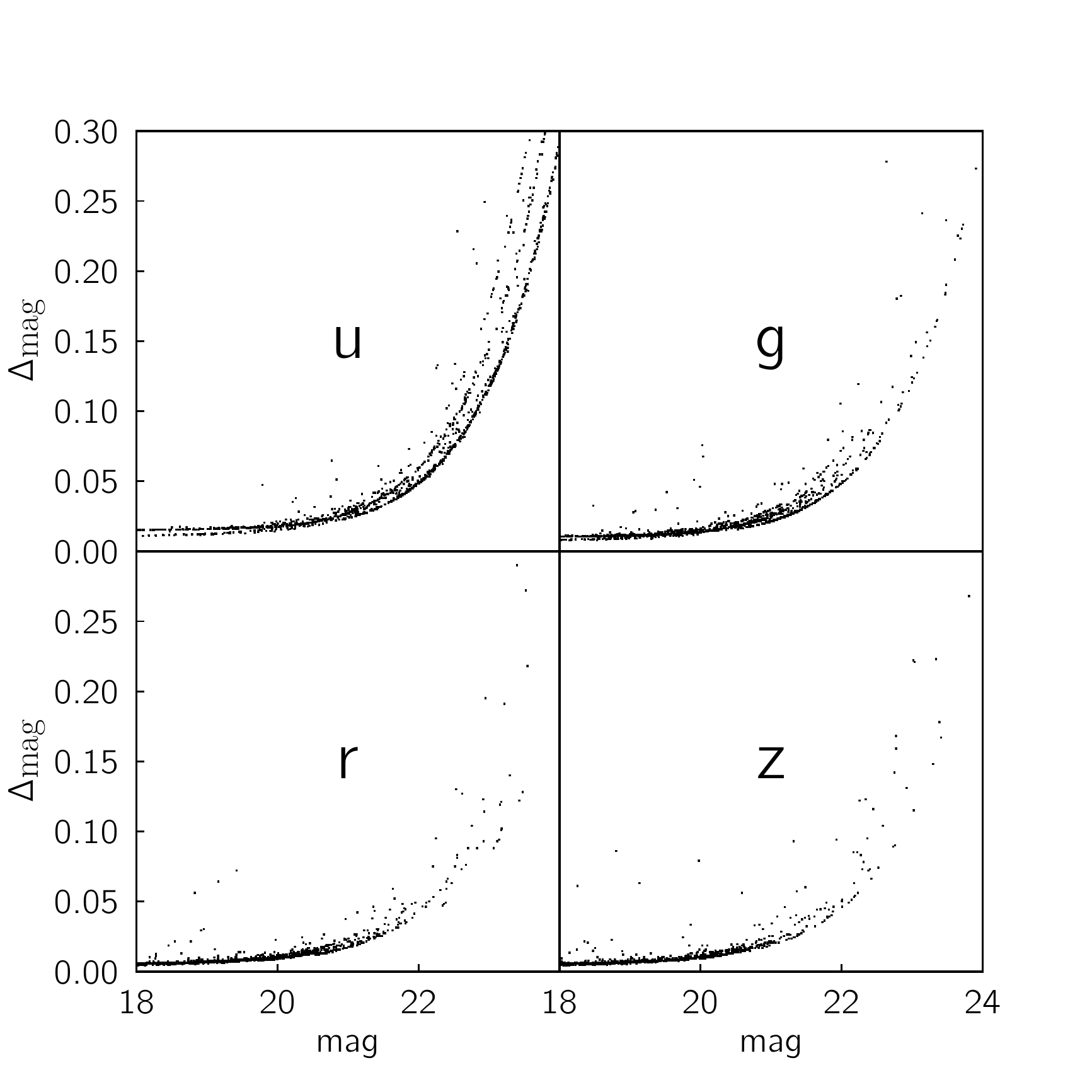}

    \caption{The $u$, $g$, $r$, and $z$ magnitudes with their respective uncertainties of the objects within our search radius of 5\arcmin. }
    \label{fig:decam_data_overview}
\end{figure}

\begin{table}
\begin{center}
\begin{tabular}{lrrrr}
\toprule
Filter &  $n_\textrm{obs}$ &  $t_\textrm{exp}$ &  FWHM &  $\sigma_\textrm{FWHM}$ \\
 - &  1  &  s &  1\arcsec &  1\arcsec \\
\midrule
u &                 5 &               400 &  1.08 &                    0.07 \\
g &                 5 &                50 &  0.97 &                    0.10 \\
r &                 5 &                50 &  0.85 &                    0.04 \\
z &                 5 &               100 &  0.76 &                    0.03 \\
\bottomrule
\end{tabular}
\end{center}
\label{tab:obs_table}
\caption{Photometry acquired with DECAM on the night of 2017 1 30}
\end{table}
The images were acquired on the night of 2017 Jan 30 with the \glsfirst{decam} instrument mounted on the 4-m Blanco telescope located at the \gls{ctio}. Table~\ref{tab:obs_table} shows the filter combination, exposure time and seeing in this night using a dither pattern of $\approx60\arcsec$. We used the standard fields sdssj0958\_0010 and sdssj1227\_0000 in our calibration.

We processed the data using an implementation of the photpipe pipeline
modified for DECam images. Photpipe is a robust pipeline used by
several time-domain surveys \citep[e.g., SuperMACHO, ESSENCE, Pan-STARRS1;
see][]{2005ApJ...634.1103R,2014ApJ...795...44R}, designed to perform
single-epoch image processing including image calibration (e.g., bias
subtraction, cross-talk corrections, flat-fielding), astrometric
calibration, warping and image co-addition,
and photometric calibration. 

%We will describe the process in detail for a set of observations in one filter (total of 5 dithered exposures with 62 individual CCD frames). 

Observations in each filter consists of total of 5 dithered exposures with 62 individual CCD frames per exposure covering the \gls{decam} field of view. We process each CCD individually, combining the 5 dither positions per CCD. We use the \gls{2mass} and the \gls{iraf} task \textsc{msccmatch} to apply a world coordinate system to the each of the frames using pre-determined distortion terms.  This coordinate system then was used to re-project each of the 62 stacks (consisting of the 5 dithered exposures each) to a common coordinate system with the software \gls{swarp} using the \textsc{lanzcos4} method. We performed DoPhot PSF photometry on each stack resulting in 62 catalogues. This was also done for the standard star fields sdssj0958\_0010 and sdssj1227\_0000. In a given standard star field image, we have about 120 -- 150 stars for
$rz$, 50 -- 80 for $g$, and 20 --40 for $u$. We convert the PS1 magnitudes into the DECam natural system AB magnitudes, following the paper by \citep{2015ApJ...815..117S}. The typical uncertainties in the zeropoint are  <0.01, <0.02, and <0.03 mags for $riz$, $g$, and $u$, respectively.

For a given filter, we propagate the zeropoints from the standard star fields to the science images, correcting for the airmass, and
averaging the zeropoints from the different standard star images. Since we have five standard star images per filter, this decreases the
uncertainty by an additional $1/\sqrt{5}$. We apply this mean zeropoint to the \sn{1006}\ image, and create a secondary photometric catalog.
Thus the systematic uncertainty in these secondary catalogs is 0.005,
0.01, and 0.015 mags for $rz$, $g$, and $u$ respectively.

Finally, the dithering strategy generated stacked images that overlapped and thus the individual catalogues cannot just be appended but need to be merged. In the merging process, we identified stars that were within $1\arcsec$ but located on different stacks and co-added those using standard uncertainty propagation rules. This process resulted in the final catalogue for one filter and was repeated for each filter-set.

The final $u, g, r, z$ catalogue was created using a similar merging process with a match distance of 1 arcsecond. The resulting catalogue (limited to our search radius of 5\arcmin-- see Section~\ref{sec:general_priors}) can be seen in Figure~\ref{fig:decam_data_overview}. The structure seen in the uncertainty is due to the dithering: Stars at the edge of each CCD were observed fewer times than stars near the middle of the CCD. The final catalogue is available at \url{https://doi.org/10.5281/zenodo.883143}.

\section{Analysis}
\label{sec:analysis}
 \begin{figure*}
	\includegraphics[width=\textwidth,clip]{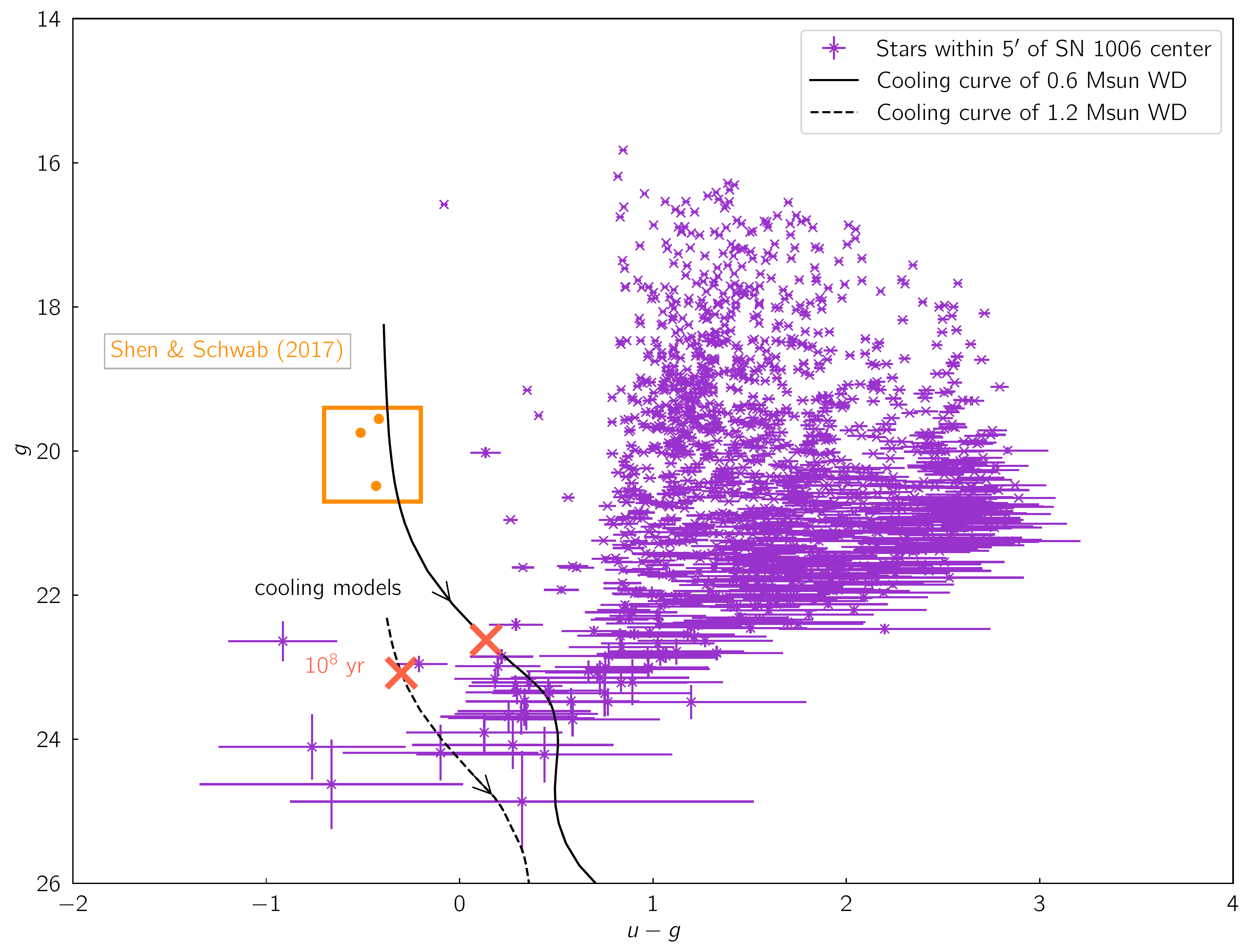}
	\caption{A CMD of all DECAM candidates within $5^{\prime}$ of the center. In addition, the fiducial blackbody models (assuming an extinction of $A_V=0.3$; see Table~\ref{tab:ks_models}) by \citet{2017ApJ...834..180S} are in orange circles. We have also added \gls{wd} cooling curves \citep[solid: 0.6 M$_\odot$, dashed: 1.2 M$_\odot$;][]{2011A&A...531L..19T} for comparison. The light red crosses on the cooling tracks mark an age of $10^8$ years.}
    \label{fig:simple_cmd}
\end{figure*}

\begin{table*}
\begin{tabular}{rrrrrrrr}
\toprule
 Mass &  Radius &  Temperature &    $u$ &    $g$ &    $r$ &    $z$ &  $u-g$ \\
 \msun &   km &  \num{1e3} K &        AB  mag &         AB  mag &          AB  mag &          AB  mag &       AB  mag \\
\midrule
  0.3 &   15000 &         45 &  19.41 &  19.78 &  20.17 &  20.78 &  -0.37 \\
  0.6 &    9100 &         50 &  20.33 &  20.71 &  21.12 &  21.73 &  -0.38 \\
  0.9 &    6600 &        150 &  19.51 &  19.97 &  20.44 &  21.11 &  -0.46 \\
\bottomrule
\end{tabular}

\caption{Magnitudes and colors of \citet{2017ApJ...834..180S}'s fiducial models at the age of SN 1006 assuming the object to be a blackbody at 2.2 kpc with extinction of $A_V=0.3$}
\label{tab:ks_models}
\end{table*}
The aim of this experiment is to identify surviving \gls{wd} companions in \sneia explosions. %We will use the scenario described in \citet[][see Table~\ref{tab:ks_models}]{2017ApJ...834..180S} as our reference scenario for surviving \glspl{wd} and will mention the other scenarios where applicable.

\subsection{Assumptions}
\begin{figure}
	% To include a figure from a file named example.*
	% Allowable file formats are eps or ps if compiling using latex
	% or pdf, png, jpg if compiling using pdflatex
	\includegraphics[width=\columnwidth]{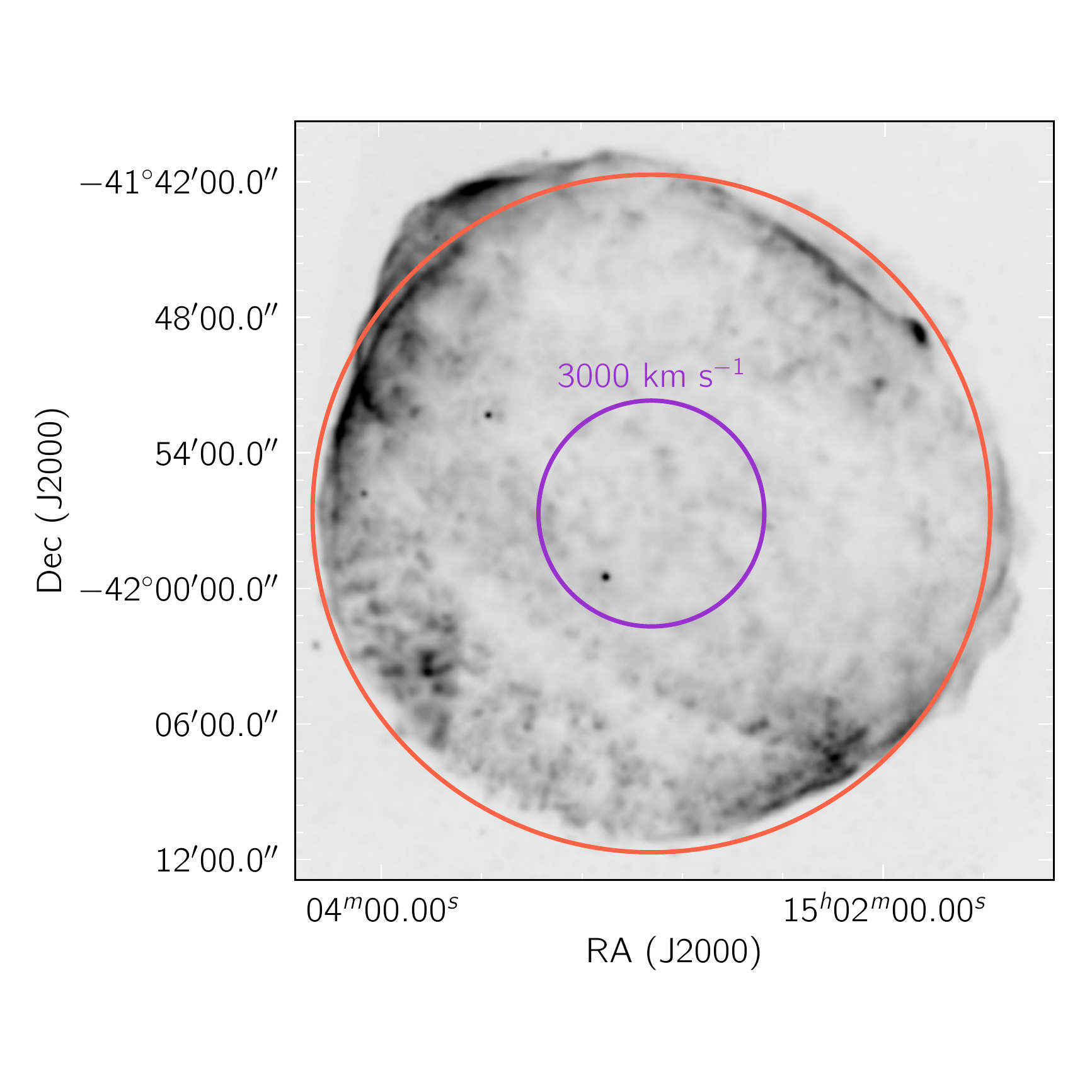}

    \caption{An overview of the \sn{1006}{} remnant in X-rays (0.5 -- 0.9 keV) de-noised using the \glsfirst{scikit-image} function \textsc{skimage.restoration.denoise\_tv\_chambolle} for better visibility. The $\approx 15\arcmin$ radius of the remnant around our chosen center is shown as the red circle.  We also show our chosen search area which corresponds to the distance moved by an object with a velocity of 3000~\kms\ in the plane of the sky.}
    \label{fig:x_ray_overview}
\end{figure}
\label{sec:general_priors}

For our analysis, we use the remnant's distance from \citet{2003ApJ...585..324W} corresponding in a Gaussian prior of $2.07 \pm 0.18$\,kpc (using a conservative value of 2.2~kpc where appropriate). 

A conservative estimate of the search area for such a \gls{wd} (including uncertainties in the center determination) is 5\arcmin\ centered around $15^\mathrm{h}02^\mathrm{m}55.4^\mathrm{s}$ $-41^\circ56{}^\prime33{}^{\prime\prime}$ \citep{2003ApJ...585..324W}. This is driven by a maximum escape velocity estimate of a potential \glspl{wd} companion presented in \citet{2017ApJ...834..180S} of 1500 -- 2000~\kms\ (corresponding to 2.4 -- 3.2\arcmin\ at a distance of 2.2~kpc). We use a simple uniform prior that assigns all stars outside of 5\arcmin\ of the center ($\approx1/3$ of the radius; see Figure~\ref{fig:x_ray_overview}) as having zero probability of being a candidate and all stars within the radius having a uniform probability.

\citet{2011ApJ...737..103S} give an extinction range estimate of $A_V =$ 0.26 -- 0.30 assuming an $R_V=3.1$, which we adopt as a uniform prior for our model. This extinction value is transformed in this work with $A_u/A_V=1.66$, $A_g/A_V=1.22$, $A_r/A_V=0.89$, and $A_z/A_V=0.48$ to the appropriate filters.

\subsection{Theoretical Surviving White Dwarf Models}

Figure~\ref{fig:simple_cmd} shows the $u-g$/g colour magnitude diagram of the candidate stars.
For a comparison of the fiducial \citet{2017ApJ...834..180S} models with the data, we approximate surviving WDs with blackbodies (using the data from Table~\ref{tab:ks_models}) and generate photometry with \gls{wsynphot}. Figure~\ref{fig:simple_cmd} shows that no viable candidates are near these models.

\subsection{White Dwarf Cooling}

We extend our search to fainter \gls{wd} candidates by allowing cooling. The \gls{wd} cooling curve \citep[][]{2006AJ....132.1221H,2011A&A...531L..19T,2011ApJ...737...28B, 2006ApJ...651L.137K} for two masses is shown in Figure~\ref{fig:simple_cmd}. Several photometry data points are consistent with old \glspl{wd} (see Figure~\ref{fig:simple_cmd}). The age $10^8$ years marked as red crosses, which is approximately when the data and models begin to overlap. In the following sections, we explore the possible parameter space allowed if a surviving white dwarf is in our data set.

\subsection{Cooling Models}

We approach the exploration of the parameter space of fainter \glspl{wd} using cooling models.  The cooling models in \citet{2011ApJ...737...28B}\footnote{\url{http://www.astro.umontreal.ca/~bergeron/CoolingModels/}} with their $u$, $g$, $r$, $z$ photometry are ideal for this task. These models span a mass range of $0.2$ -- $1.2$~\msun\ with cooling tracks ending at a little more than a few Gyr. We use interpolation (using \textsc{scipy.interpolate.CloughTocher2DInterpolator}) to efficiently explore the parameter space and obtain photometry for different ages and masses.  Specifically, we use the base-10 logarithm for the interpolation of masses and ages (as the magnitudes behave more linearly in the logarithmic space). Finally, we added a distance modulus and extinction \citep{2011ApJ...737..103S} to the interpolated magnitudes resulting in the model $M_\textrm{WD}(\textrm{mass}, \textrm{age}, d, A_V) \mapsto u, g, r, z$.

\subsection{Posterior probability}
\label{sec:posterior}

We explore all possible white dwarf parameters under the assumption that the cooled white dwarf is in our sample. Here we develop the likelihood of any candidate star being a white dwarf of a certain $\textrm{mass}, \textrm{age}, d, A_V$ given the photometric data.

Given a single star and assuming it is a cooling white dwarf with given parameters $\textrm{mass}, \textrm{age}, d, A_V$, the likelihood to observe its photometric data $D_j=(u_j, g_j, r_j, z_j)$ is:

\begin{align}
P(D_j|\textrm{mass}, \textrm{age}, d, A_V) = \mathcal{L}_j \propto \exp\left(-\chi_j^2/2\right)
\end{align}
with
\begin{align}
\chi_j^2 = \left(\frac{u_{j, \textrm{data}} - u_\textrm{model}}{\sigma_u}\right)^2 + \ldots 
+ \left(\frac{z_{j, \textrm{data}} - z_\textrm{model}}{\sigma_z}\right)^2
\end{align}

We now assume that exactly one of the stars in our sample is the white dwarf. Not knowing which one it is, 
and assuming each is equally probable a priori, we therefore sum their individual probabilities:

% J.B.: Note that in principle here you could put in a distance weighting; but probably unnecessarily complex

\begin{align}
P(D|\textrm{mass}, \textrm{age}, d, A_V) = \sum_{j=0}^N P(D_j|\textrm{mass}, \textrm{age}, d, A_V)
\end{align}

%For a set of $N$ stars the likelihood that this contains stars that are consistent with a certain combination of mass, age, $d$, and $A_V$ is the sum over all of their individual likelihoods:

Following Bayes theorem, to compute parameter probabilities from this likelihood, we need to multiply it with the parameter space priors:

\begin{align}
P(\textrm{mass}, \textrm{age}, d, A_V|D) \propto& P(D|\textrm{mass}, \textrm{age}, d, A_V) \times\\
\nonumber & P(\textrm{mass}, \textrm{age}, d, A_V) 
\end{align}

We use the priors of Section~\ref{sec:general_priors}. We additionally assume log-uniform prior for the \gls{wd} age between $10^6$ -- $10^{10}$ years and a uniform prior for the mass between $0.2\msun$ -- $1.2\msun$. 

%[\textbf{T.Do:} - This is confusing to me -- why are you summing over all the candidates? I would have thought that you would do this for each individual candidates and look for the likelihood that they could be at the right distance and age?] 

%Our main aim in this work is to study the probability of the existence of one candidate within our \gls{decam} dataset given the model parameters. This means that we can approximate the sum over all \emph{N} candidates by just using the most likely candidate for each combination of mass, age, $d$, and $A_V$:

%\begin{align}
%&P(u, g, r, i|\textrm{mass}, \textrm{age}, d, A_V) \approx \\
% \nonumber \approx  \max( &P(u, g, r, z|\textrm{mass}, \textrm{age}, d, A_V)_0, \ldots\\
%\nonumber, &P(u, g, r, z|\textrm{mass}, \textrm{age}, d, A_V)_N)
%\end{align}
%[\textbf{T.Do:} This seems inconsistent with equation 2. In that one, you are already adding all the likelihoods together... ]

\subsection{Parameter Space Exploration}

\begin{figure*}
	% To include a figure from a file named example.*
	% Allowable file formats are eps or ps if compiling using latex
	% or pdf, png, jpg if compiling using pdflatex
	\includegraphics[width=\textwidth]{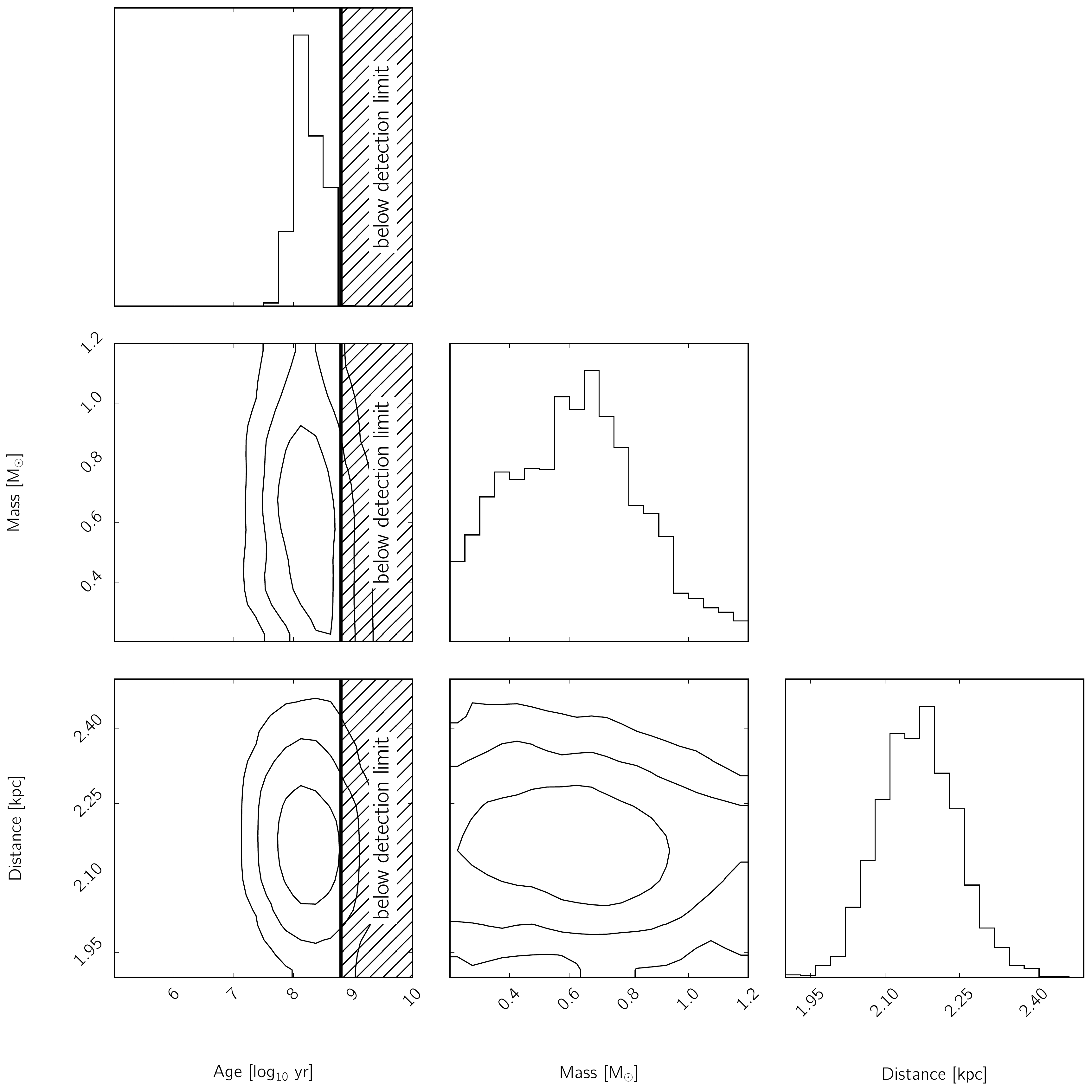}

    \caption{Posterior WD properties given the likelihood and priors in Section~\ref{sec:posterior} sampled using \textsc{nestle}.  We marginalize in these collection of plots over $A_V$. Note that ages above ${10}^8$ years are preferred.}
    \label{fig:corner_plot_age_mass} 
\end{figure*}

\begin{figure*}
	% To include a figure from a file named example.*
	% Allowable file formats are eps or ps if compiling using latex
	% or pdf, png, jpg if compiling using pdflatex
	\includegraphics[width=\textwidth]{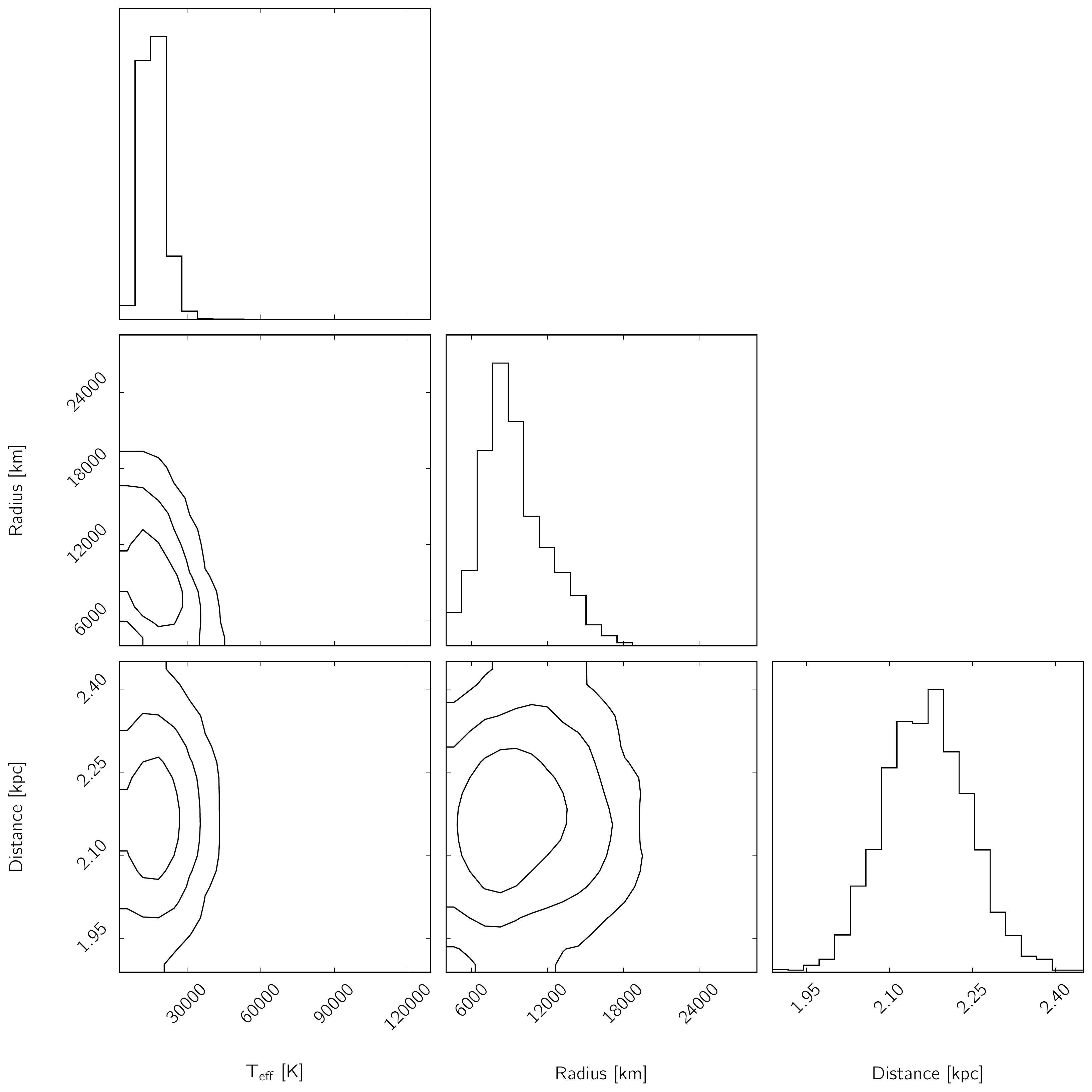}

    \caption{Exploration of the parameter space as given in Figure~\ref{fig:corner_plot_age_mass} but using a conversion of the age and mass to radius of temperature by interpolating on the \citet{2011ApJ...737...28B} grid. We marginalise in these collection of plots over $A_V$}
    \label{fig:corner_plot_temperature_radius}
\end{figure*}

The parameter space is sampled using the \gls{multinest} algorithm and using the implementation available at \url{https://github.com/kbarbary/nestle}. Figure~\ref{fig:corner_plot_age_mass} shows the exploration of the posterior probability including three confidence intervals (68\%, 95\%, and 99\%). We have also given an approximate region where the \gls{wd} with the given parameters would be close or below the detection limit of our data. Figure~\ref{fig:corner_plot_temperature_radius} shows the sample converted from mass and age to temperature and radius (using interpolation). In this figure, we have not marked the areas that are below the detection limit as it is not as straight-forward as with the mass-age parameter space.

\section{Discussion}
\label{sec:results}

We have placed stringent limits on a surviving companion to \sn{1006}. We have excluded all young white dwarf models ($\lesssim 10^8$ yr) based on a comparison between DECam photometry and theoretical white dwarf models.
Previous shallower searches of \citet{2012Natur.489..533G, 2012ApJ...759....7K} have already ruled out red-giant and main-sequence star companions.

We have compared our photometric data with two methods: Figure~\ref{fig:simple_cmd} shows a traditional visualization of photometric data. Figure~\ref{fig:corner_plot_age_mass} and Figure~\ref{fig:corner_plot_temperature_radius} show the posterior probability for any surviving \gls{wd} existing in the data. The posterior probability for the \gls{wd} cooling models disallows relatively young \glspl{wd} with ages of $\lesssim10^8$ years. At low masses, the cooling curve (solid line in Figure~\ref{fig:simple_cmd}) crosses several of our candidate stars at these ages, while at higher masses (dashed line) we have fewer objects. In the probability distribution this makes high masses slightly preferred. The posterior distance probability density distribution is the same as the prior probability.

%[\textbf{T.Do:} I'm having trouble understanding this. My confusion probably has to do with how the probabilities are combined. It is simpler for me to think about the candidates with the highest probability of being a companion and seeing how those differ.]

%[\textbf{J.B:} Also not clear to me. I think you are trying to explain why cooling does not bring high-mass  WD closer to the data. Also, "seemingly" does not fit here.]

\subsection{Close surviving companion WDs}

Double \gls{wd} binaries have been studied as possible \gls{snia} progenitor systems in various incarnations for decades \citep{1984ApJS...54..335I,1984ApJ...277..355W}.  Recent work has raised the possibility that the high temperatures and densities reached during double WD mergers involving a helium or low-mass C/O WD can trigger a helium-powered detonation on the surface of the more massive WD \citep{2010ApJ...709L..64G,2012ApJ...746...62R,2013ApJ...770L...8P}.  This helium shell detonation can then lead to a carbon core detonation and subsequent SN Ia in a variant of the classic double detonation scenario \citep{1982ApJ...257..780N,1990ApJ...354L..53L,2014ApJ...785...61S}.

The importance of including the necessary isotopes and reaction rates in simulations of helium shell detonations was pointed out by \citet{2014ApJ...797...46S}, who found that such detonations could be triggered in much smaller helium layers than previously realized.  This led to the possibility that double detonation \sneia could occur early enough in the the double WD merging process that the less massive WD is not completely tidally disrupted when the SN Ia occurs: that is, the less massive companion WD may survive the explosion of the more massive WD.

Double detonations in double WD systems have also been proposed in stably mass transferring AM Canum Venaticorum binaries \citep{2007ApJ...662L..95B,2007A&A...476.1133F,2009ApJ...699.1365S,2010A&A...514A..53F}. For extreme mass ratios, these systems may avoid unstable mass transfer (although see \citealt{2015ApJ...805L...6S}) and lead to accreted heliums shells of $\sim 0.01 \ M_\odot$ on the more massive WD. Convection in these shells becomes inefficient and may trigger a helium detonation, which then sets off the core detonation and subsequent \snia.  The donor \gls{wd} in these systems will remain undisrupted because it was undergoing stable Roche lobe overflow.

Thus, there are several avenues that may lead to a close-in surviving companion WD following a SN Ia.  \citet{2017ApJ...834..180S} explored the effect of the $^{56}$Ni that is captured from the SN Ia ejecta by nearby surviving WDs.  The high temperature and complete ionization of the captured nuclei results in a strongly suppressed $^{56}$Ni decay rate, yielding a long-lived luminous outflowing wind.  Depending on the mass of the surviving companion and the amount of $^{56}$Ni that is captured, the companion may be visible long after the SN Ia occurred.

The orange circles in Figure \ref{fig:simple_cmd} show the expected $g$ magnitudes and $u-g$ colours of \citet{2017ApJ...834..180S}'s fiducial $0.3$, $0.6$, and $0.9 \ M_\odot$ surviving companion \glspl{wd} at the present age of SN 1006.  It is clear that none of the stars within $5'$ of the remnant's center have the correct magnitude and color to match \citet{2017ApJ...834..180S}'s fiducial models.  Thus, these models have been ruled out.

However, important caveats remain regarding the appearance of a surviving companion WD.  The models of \citet{2017ApJ...834..180S} used ad hoc estimates for the amount and thermodynamic conditions of the captured $^{56}$Ni mass; in particular, the mass of the captured $^{56}$Ni and thus the luminosity should be regarded as an upper limit.  A better quantitative estimate of the initial conditions for the outflowing wind requires hydrodynamical simulations that have yet to be undertaken.  Furthermore, the models assume a constant opacity of $0.2 {\rm \, cm^2 \, g^{-1}}$ in the $^{56}$Ni-rich layer; this is likely a strong underestimate of the true opacity due to iron-group line blanketing.  Higher opacities will also significantly alter the luminosity predictions by both increasing the mass outflow rate and changing the colors. An exploration of these effects awaits future simulations; until then, we cannot strongly rule out the presence of a surviving companion WD initially near the explosion site.

\subsection{Spin-up/Spin-down model}

\citet{2011ApJ...730L..34J} and \citet{2011ApJ...738L...1D} propose a single-degenerate \snia scenario in which the angular momentum of the accreted material spins up the WD, allowing it to grow above the canonical (non-rotating) critical mass needed for explosion $(\approx 1.38\msun)$.  The donor star continues to evolve and in many cases also becomes a WD.  Since mass transfer has ceased, the accretor WD spins down and the central density increases, until eventually the explosion is triggered.

The mechanisms of angular momentum loss and redistribution that would operate are uncertain, and thus the delay timescale between the end of mass transfer and the explosion is unknown. \citet{2008ApJ...679..616P} and \citet{2017A&A...602A..55N} find that the WD should be close to solid body rotation during the accretion phase.  This implies that one would not form the highly super-Chandrasekhar objects allowed in the presence of differential rotation and suggests the spin-down timescale is not associated with the internal redistribution of angular momentum, but rather the timescale for the loss of angular momentum from the system.

The requirement that the outcome be a normal \sneia\ may itself impose some timescale constraints.  For their rotating WD models, \citet{2005A&A...435..967Y} suggest that spin-down timescales of $>10^6$ years imply central densities such that the explosion would violate nucleosynthetic constraints \citep{1999ApJS..125..439I}.  \citet{1991ApJ...367L..19N} find that once the accretor WD has crystallized $(\ga 3\times10^{9}$ years), carbon ignition leads instead to accretion-induced collapse.

The data and analysis presented in this paper (see Figure~\ref{fig:corner_plot_age_mass}) show all possible WDs in \sn{1006}{} are at a minimum a $\approx 10^8$ years old.  This provides a significant observational constraint on spin-up/spin-down models as an explanation for ``prompt'' \sneia or for systems that show interaction with material that must have been produced in the relatively recent past (e.g., nova shells).

\section{Conclusions}
\label{sec:conclusion}

We present very deep multi-color ($u, g, r, z$) photometric data of stars in the center of the \sn{1006}{} remnant in search of surviving WDs. The data show no bright unambiguously identifiable \glspl{wd} and thus this data is inconsistent with many spin up/down models as well as model presented in \cite{2017ApJ...834..180S}. This suggests various hypotheses that are consistent with the data. The first hypothesis is that a surviving \gls{wd} exists but is not as bright as predicted, due to either incorrect assumptions or a limited exploration of parameter space. The majority of red faint objects near the \gls{wd} cooling curves will likely be foreground stars, but too faint to be detected by Gaia. Unfortunately, many of the other remnants that could be used to test this hypothesis are heavily affected by extinction. This extinction is very detrimental in using the $u-g$-color to distinguish \glspl{wd} from unrelated faint fore/back-ground objects. The second hypothesis is that these surviving \gls{wd} companions do not exist. This hypothesis could be extended (due to the various non-detections of companions in literature) to the claim that generally \sneia do not leave any survivors. This would firmly point to the merger and complete disruption of \glspl{wd}. However, the merger hypothesis has for now no easily falsifiable predictions except for the detection of a companion. 

\section*{Acknowledgements}

W.~E.~Kerzendorf was supported by an ESO Fellowship. Support for this work was provided by NASA through Hubble Fellowship grant \# HST-HF2-51382.001-A awarded by the Space Telescope Science Institute, which is operated by the Association of Universities for Research in Astronomy, Inc., for NASA, under contract NAS5-26555.  KJS was supported by grants from the NASA Astrophysics Theory Program (NNX15AB16G and NNX17AG28G).

In addition to the software packages mentioned in the paper, we used the following software \glsfirst{astropy}, \glsfirst{numpy} \glsfirst{scipy}, \glsfirst{pandas}, \gls{matplotlib} and \glsfirst{aplpy} to analyze and visualize the data. This research has made use of NASA's Astrophysics Data System Bibliographic Services as well as the DeepThought literature discovery tool \citep{2017arXiv170505840K}.

%%%%%%%%%%%%%%%%%%%%%%%%%%%%%%%%%%%%%%%%%%%%%%%%%%

%%%%%%%%%%%%%%%%%%%% REFERENCES %%%%%%%%%%%%%%%%%%

% The best way to enter references is to use BibTeX:

\bibliographystyle{mnras}
\bibliography{wekerzendorf} % if your bibtex file is called example.bib

% Alternatively you could enter them by hand, like this:
% This method is tedious and prone to error if you have lots of references

%%%%%%%%%%%%%%%%%%%%%%%%%%%%%%%%%%%%%%%%%%%%%%%%%%

%%%%%%%%%%%%%%%%% APPENDICES %%%%%%%%%%%%%%%%%%%%%

%%%%%%%%%%%%%%%%%%%%%%%%%%%%%%%%%%%%%%%%%%%%%%%%%%

% Don't change these lines
\bsp	% typesetting comment
\label{lastpage}
\end{document}